\documentclass[twocolumn,showpacs,preprintnumbers,amsmath,amssymb]{revtex4}
\usepackage{graphicx}% Include figure files
\usepackage{dcolumn}% Align table columns on decimal point
\usepackage{bm}% bold math

\newcommand{\Hb}{\bf}

\begin{document}

\preprint{APS/123-QED}

\title{Scattering of backward spin waves in a one-dimensional magnonic crystal}

\author{A.V.~Chumak}

 \email{chumak@physik.uni-kl.de}
 \altaffiliation[\\ Also at ]{Taras Shevchenko National University of Kiev, Ukraine.}

\author{A.A.~Serga}

\author{B.~Hillebrands}

\affiliation{Fachbereich Physik  and Landesforschungszentrum OPTIMAS, Technische Universit\"at Kaiserslautern,
67663 Kaiserslautern, Germany}

\author{M.P.~Kostylev}
\affiliation{School of Physics, University of Western Australia, Crawley, Western Australia 6009,
Australia}

\date{\today}

\begin{abstract}
Scattering of backward volume magnetostatic spin waves from a one-dimensional magnonic crystal, realized by a
grating of shallow grooves etched into the surface of an yttrium-iron garnet film, was experimentally studied.
Rejection frequency bands were clearly observed. The rejection efficiency and the frequency width of the
rejection bands increase with increasing groove depth. A theoretical model based on the analogy of a spin-wave
film-waveguide with a microwave transmission line was used to interpret the obtained experimental results.
\end{abstract}

\pacs{75.50.Gg, 75.30.Ds, 75.40.Gb}
% PACS, the Physics and Astronomy Classification Scheme.

\maketitle

Artificial media with periodic lateral variation of their magnetic properties, so-called magnonic crystals, are
the magnetic analogue to photonic and sonic crystals and are suitable for operation in the microwave frequency
range. Spectra of spin-wave excitations in magnonic crystals are considerably modified with respect to uniform
media and exhibit features such as full band gaps \cite{BH Kolodin}, where spin waves are not allowed to
propagate. Due to the wide tunability of their characteristic properties magnonic crystals offer excellent
potential for the investigation of linear and nonlinear spin-wave dynamics \cite{MC review}.

Both ferromagnetic and ferrite \cite{MC review} media can be artificially structured for fabrication of magnonic
crystals of different dimensionality.

We have chosen a single-crystal ferrite film of yttrium iron garnet
(YIG) as the magnetic material because of the extremely small
magnetic relaxation. Furthermore, we have focused on one-dimensional
magnonic crystals because they allow for operation at only one
eigenmode of the film. The latter is determined by the angle between
the wave propagation direction and the external magnetic field
orientation. Previously, only the surface magnetostatic waves (wave
propagation direction is perpendicular to the magnetic field applied
in the film plane) and forward magnetostatic waves (external field
is oriented perpendicular to the film plane) were studied in
magnonic crystals \cite{MC review, MC skyes, MC MSSW, MC FVMSW}.
However, nowadays special attention is focused on spin waves of
backward type, so-called backward volume magnetostatic waves
(BVMSW), due to their special benefits both for research on
nonlinear spin-wave dynamics \cite{bulet, reversal momentum} and
application \cite{convolver, logic}. Here we present experimental
and theoretical results on scattering of BVMSW from a structure with
periodic changes of the film thickness.

To fabricate the magnonic crystal a 5.5\,$\mu$m-thick YIG film was
used which was epitaxially grown along the (111) crystallographic
axis. Hot orthophosphoric acid etching and photolithography was used
to prepare the grooves. The lithography was based on a standard
photoresist AZ~5214E baked by UV irradiation which is stable against
hot 160 $\mathrm{^o}$C orthophosphoric acid. The mask had $N = 20$
parallel lines $w = 30$\,$\mu$m in width and spaced 270\,$\mu$m from
each other, so the lattice constant was $a = 300\,\mu$m. The groves
were perpendicularly oriented with respect to the spin-wave
propagation direction. In order to study the dependence of crystal
characteristics on the groove depth $\delta$ the grooves were etched
in 100\,nm steps from 100\,nm to 1\,$\mu$m. The grooves depth was
controlled by the etching time and measured using a profilometer.
Anisotropic chemical etching caused by the YIG crystallographic
structure was observed: the speed of etching parallel to the film
plane was approximately ten times larger than in the perpendicular
direction, so the final groove depth profile along the direction of
wave propagation had a trapezoidal shape.

\begin{figure}
\includegraphics[width=0.9\columnwidth]{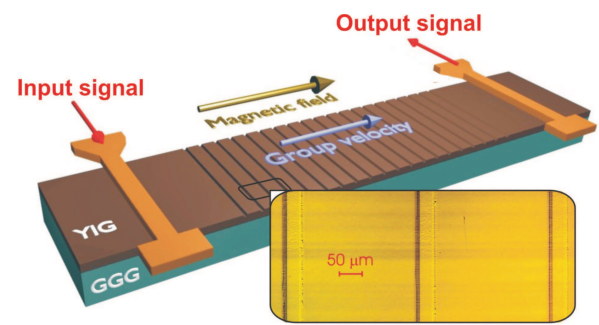}
\caption{\label{fig:epsart1} Sketch of magnonic crystal structure used in the experiments.}
\end{figure}

Two microstrip antennae placed 8\,mm apart from each other on each side of the magnonic crystal area were used to
excite and receive the dipolar spin waves, see Fig.~1. A bias magnetic field of 1845\,Oe was applied in the plane
of the YIG film stripe along its length and parallel to the direction of spin-wave propagation. Thus the
conditions for backward volume magnetostatic wave (BVMSW) propagation are given. The microwave signal power was
equal to 1\,mW to avoid any non-linear processes.

\begin{figure}
\includegraphics[width=0.95\columnwidth]{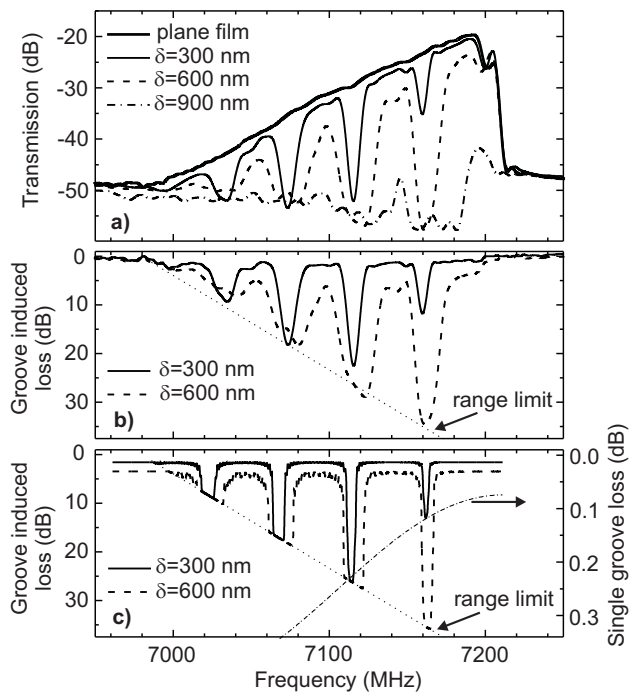}
\caption{\label{fig:epsart2} (a) - BVMSW microwave transmission characteristics for an unstructured film
(bold line) and for a magnonic crystals for different groove depths $\delta$. (b) - experimentally
measured loss induced by the magnonic crystal structure. (c) - calculated loss. Parameters of
calculation: groove number $N = 20$, groove width on bottom $w = 30$\,$\mu$m, lattice constant $a =
300\,\mu$m, film thickness $d_\mathrm{0} = 5.5$\,$\mu$m, saturation magnetization $4 \pi M_\mathrm{0} =
1750$\,G, bias magnetic field $H_\mathrm{0} = 1845$\,Oe, efficiency coefficient $\eta=5$, resonance line
width $\Delta H = 0.5$\,Oe, surface damage coefficient $\zeta = 30$. The dash-dotted line shows the
calculated loss induced by a single groove of 300\,nm in depth.}
\end{figure}

The experimental BVMSW transmission characteristics for the unstructured film as well as for magnonic
crystals with $\delta$~=~300, 600 and 900\,nm are shown in Fig.~2(a). Figure~2(b) demonstrates groove
structure induced losses calculated as a difference between the transmission characteristics for the
unstructured and structured films. The dotted straight line indicates the limit of the dynamic range of
the experimental setup.

From Fig.~2(b) one sees that the grooves as shallow as 300\,nm
result in the appearance of a set of rejection bands (or
transmission gaps), where spin-wave transmission is highly reduced.
According to the condition for Bragg reflection, higher-order
rejection bands correspond to larger spin-wave wave numbers. In the
case of BVMSW the latter corresponds to lower frequencies. From the
depths and the frequency widths $\Delta{f}$ of the gaps one sees
that the efficiency of the rejection increases with increase in the
order of Bragg reflection. This suggests that BVMSW with smaller
wavelengths are more sensitive to the introduced inhomogeneities.

Both Fig.~2(a) and 2(b) demonstrate that an increase in $\delta$ leads to an increase in the rejection efficiency
and in $\Delta{f}$. A small frequency shift of the minima of transmission towards higher frequencies is observed,
as well as an increase in insertion losses in the transmission (i.e. allowed) bands. For $\delta = 900$\,nm the
insertion loss in the whole spin-wave band is so important that almost no spin-wave propagation is observed (see
Fig.~2(a)).

The points in Fig.~3 represent the insertion loss (upper panel) and the frequency width $\Delta{f}_1$
measured for the first transmission gap (the central frequency of the gap is 7160\,MHz). This rejection
band was chosen because of its largest experimental dynamic range. From Fig.~3(a) one sees that the
difference in insertion loss for transmission and rejection bands can reach 30\,dB for
$\delta$=0.5\,$\mu$m. With increase in the groove depth the rejected power exceeds the dynamic range of
the experimental setup, and the transmission loss cannot be measured. The observed frequency shift of
the minimum of transmission with increase in $\delta$ was 10-20\,MHz. The maximum rejection band width
$\Delta{f}_1$ was almost 50\,MHz.

A theoretical description of the scattering of dipole spin waves from inhomogeneities is given by a singular
integral equation
%\begin{equation}
${\bf m}({\bf r})  =  4 \pi \hat{\kappa}(f) \cdot \left( \int_{V}\hat{G}({\bf r}-{\bf r}') {\bf m}({\bf
r'})d{\bf r}'+{\bf h}({\bf r})\right)$,
%\end{equation}
where $\bf{m}$ is the dynamic magnetization, ${\bf h}$ is the microwave field of the input antenna,
$\hat{\kappa}(f)$ is the tensor of microwave magnetic susceptibility which depends on the spin wave
carrier frequency $f$ and the magnetic parameters of the film, $\hat{G}({\bf r}-{\bf r}')$ is the
Green's function of dipole magnetic field, and $\bf{r}'$ is the radius vector. The integration is over
the volume of the magnetic structure, thus the inhomogeneity of film thickness is taken into account.
The quasi-1D dipole field \cite{guslienko} of the lowest BVMSW thickness mode decays at the distance of
a few film thicknesses. Therefore, as the width of the grooves $w$ is much smaller than $a$, most of the
distance between the consecutive grooves the spin wave travels as an eigenmode of a continuous film of
thickness $d_0$. For these sections the integral equation reduces to a simple formula (see Eq.~(50) in
\cite{kostylev}) which shows that between the grooves the transmitted and reflected waves only
accumulate phase and decay (due to intrinsic magnetic damping). Thus, in order to describe formation of
stop bands one has to consider scattering of a BVMSW from just one groove. The effect of a set of the
grooves is obtained by cascading the structure periods using matrices of scattering T-parameters {\Hb}
and taking interference effects into account.

Therefore in this work instead of solving the singular integral equation we considered the magnonic
crystal as a periodical sequence of sections of regular transmission lines with different propagation
constants (different spin-wave wave numbers) for the same carrier frequency. We neglect the fact that
the groove edges are oblique and consider the groove cross-section as a rectangle with the same depth
and having the same area. The T-matrix $T^{(1)}$ for a section of unstructured film of the length $a-w$
has diagonal components only: $\mathrm{T}^{(1)}_{11} = \mathrm{T^{(1)\ast}_{22}} = e^{(-ik+k''_0)
(a-w)}$, where $k$ is the spin-wave wavenumber in the unstructured film, $k''_0 = \gamma \Delta H /(2
v_\mathrm{gr})$ is the rate of spin-wave spatial damping, $\gamma$ is the gyromagnetic ratio, $\Delta H$
is the ferromagnetic resonance linewidth, and $v_\mathrm{gr}$ is the spin-wave group velocity.
Similarly, the T-matrix $T^{(3)}$ for a regular spin-wave film waveguide with the thickness
$d=d_0-\delta$ is $\mathrm{T^{(3)}_{11}} = \mathrm{T^{(3)\ast}_{22}} = e^{(-ik + k''_\mathrm{g})w
d_\mathrm{0} / d}$; $\mathrm{T^{(3)}_{12}} = \mathrm{T^{(3)}_{21}} = 0$, where $k''_\mathrm{g}$ is the
spin-wave damping rate for the groove. Here we used the fact that the BVMSW dispersion law for small
wave numbers $kd \ll 1$ is practically linear, therefore the spin-wave wave number in the grooves is $k
d_\mathrm{0}/d$. To describe the increase in the loss in the pass bands with increase in the groove
depth we introduce an empirical parameter $\zeta$ which accounts for larger contribution of two-magnon
scattering processes in the areas which underwent anisotropic etching \cite{etching}. Then the damping
rate in the grooves can be expressed as $k''_\mathrm{g} = k''_0 (1+\zeta \delta/d_0)$.

\begin{figure}
\includegraphics[width=0.9\columnwidth]{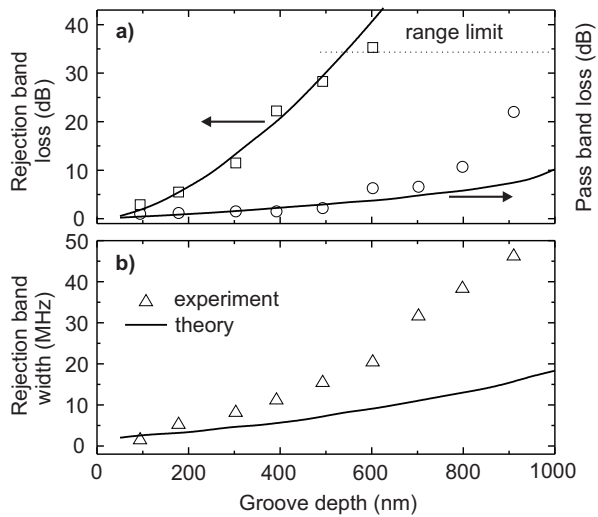}
\caption{\label{fig:epsart3} (a) - insertion loss in rejection band (squares) and pass band (circles) a function
of the groove depth; solid lines show corresponding calculated dependencies. (b) - experimental (triangles) and
theoretical (line) first rejection band width as a function of grooves depth.}
\end{figure}

There are reflections from the junctions of the consecutive sections. The T-matrix for the front edge of
the groove is $T^{(2)}$ and that for the rear edge is $T^{(4)}$. Therefore the T-matrix for one period
of the structure is, as follows: $\mathrm{T} = [\mathrm{T^{(1)}} \cdot \mathrm{T^{(2)}}\cdot
\mathrm{T^{(3)}} \cdot \mathrm{T^{(4)}}]$. To obtain the T-matrix for the whole groove sequence on has
to raise $\mathrm{T}$ to the ${N}$-th power. The reflection coefficient for a junction of two sections
of regular waveguides is $\Gamma$. Following \cite{transm line}, the transmission coefficient through
the junction is $1-\Gamma$. Then one obtains: $\mathrm{T^{(2)}_{11}} = \mathrm{T^{(2)}_{22}}
 = (1-\Gamma)^\mathrm{-1}$; $\mathrm{T^{(2)}_{12}} =
\mathrm{T^{(2)}_{21}} = \Gamma (1-\Gamma)^\mathrm{-1},$
and
$\mathrm{T^{(4)}_{11}} = \mathrm{T^{(4)}_{22}}
 = (1+\Gamma)^\mathrm{-1}$; $\mathrm{T^{(4)}_{12}} =
\mathrm{T^{(4)}_{21}} = -\Gamma (1+\Gamma)^\mathrm{-1}$.

Now one has to specify the form of the reflection coefficient $\Gamma$. Here we use the analogy of the
change in the film waveguiding properties to a change of characteristic impedance $Z$ of a microwave
transmission line \cite{kalinikos}. We assume that the change of the characteristic impedance of a
spin-wave waveguide because of change of YIG-film thickness is due to change of the film effective
inductance. Then the characteristic impedance is linearly proportional to the propagation constant (to
the spin-wave wave number in our case), and from Eq.~(3) in \cite{transm line} we arrive at a formula
for the reflection coefficient for the wave incident onto the edge of the groove from the unstructured
section of the film: $\Gamma = \eta {\delta}/({2d_0-\delta})$.

For the wave incident onto the same junction in the reverse direction $\Gamma_{-}=-\Gamma$ which has
been already taken into account in the expressions for the T-matrices above. The phenomenological
parameter $\eta > 1$ is introduced in this formula to account for eventual factors not taken into
account in this simplistic model.

The results of our numerical computation of $\mathrm{T}$ in the 20-th power are shown in Fig.~2(c) and Fig.~3.
One sees that this model is in qualitative  agreement with all the tendencies we see in the experiment. In
particular, it shows the observed increase in the rejection efficiency with increase in $k$, and the correct
behavior of all characteristics of the rejection bands as functions of groove depth $\delta$.

The formation of transmission gaps is due to multiple reflections from edges of all grooves which form
partial standing waves in the space between the grooves. The strength of these reflections is entirely
determined by transmission and reflection of a single groove. The latter is shown (in a suitable scale)
in Fig.~2(c) for $\delta$= 300\,nm by a dash-dotted continuous line. One sees that the peaks of
rejection for the 20-grove structure follow this line. However, the quantitative agreement of this model
is poor, unless one introduces a value of $\eta$ considerably larger than 1 (in our calculation
$\eta=5$). In particular, for $\eta=1$, when the reflection from a grove edge is entirely due to
transformation of wavenumber, the model considerably underestimates the depth of transmission gaps. This
suggests that other effects such as transformation of modal distribution of dynamic magnetization and
eventual generation of other thickness modes (and, possibly, back-transformation) through 2-wave
scattering processes give significant contributions to $\Gamma$.

In conclusion, in this work we experimentally demonstrated that in the BVMSW configuration a one-dimensional
magnonic crystal showed excellent spin-wave signal rejection of more than 30\,dB. The width of the rejection
bands exceeded the values for the other spin-wave configurations \cite{MC review, MC skyes, MC MSSW, MC FVMSW},
and could be controlled by the groove depth. A simple model was proposed which is in qualitative agreement with
the experimental results.

Financial support by the DFG SE 1771/1-1, Australian Research Council, and the University of Western Australia is
acknowledged. Special acknowledgments to Dr. Sandra Wolff and the Nano+Bio Center, TU Kaisers\-lautern.

\end{document}